# A stabilized 18 GHz chip-scale optical frequency comb at $2.8\times10^{-16}$ relative inaccuracy

S.-W. Huang[1,*], J. Yang[1], M. Yu[2], B. H. McGuyer[3], D.-L. Kwong[2], T. Zelevinsky[3], and C. W. Wong[1,*]

[1]*Mesoscopic Optics and Quantum Electronics Laboratory, University of California, Los Angeles, CA 90095*

[2] *Institute of Microelectronics, A*STAR, Singapore 117865*

[3]*Department of Physics, Columbia University, New York, NY 10027*

*Author e-mail address: swhuang@seas.ucla.edu, cheewei.wong@ucla.edu

**Optical frequency combs, coherent light sources that connect optical frequencies with microwave oscillations, have become the enabling tool for precision spectroscopy, optical clockwork and attosecond physics over the past decades. Current benchmark systems are self-referenced femtosecond mode-locked lasers, but four-wave-mixing in high-$Q$ resonators have emerged as alternative platforms. Here we report the generation and full stabilization of CMOS-compatible optical frequency combs. The spiral microcomb's two degrees-of-freedom, one of the comb line and the native 18 GHz comb spacing, are first simultaneously phase-locked to known optical and microwave references. Second, with pump power control, active comb spacing stabilization improves the long-term stability by six orders-of-magnitude, reaching an instrument-limited 3.6 mHz/$\sqrt{\tau}$ residual instability. Third, referencing thirty-three of the nitride frequency comb lines against a fiber comb, we demonstrate the comb tooth-to-tooth frequency relative inaccuracy down to 53 mHz and $2.8\times10^{-16}$, heralding unprecedented chip-scale applications in precision spectroscopy, coherent communications, and astronomical spectrography.**

Recently continuous-wave (cw) pumped high-$Q$ microresonators have emerged as promising alternative platforms for ultrashort pulse and optical frequency comb generation [1-10]. Microresonator-based optical frequency combs, or Kerr frequency combs, are unique in their compact footprints and suitably large comb spacings, thereby expanding the already remarkable applications of optical frequency combs. Microresonators with microwave free spectral ranges have recently been advanced in both whispering gallery mode (WGM) structures [11-14] and planar ring geometries [15-17]. Even though CMOS-compatible ring resonators are particularly attractive because of the monolithic electronic and photonic integration capabilities, until now there has been no demonstration nor evidence of full stabilization and precision uniformity of chip-scale planar microresonators, furthering from a prior silica microtoroid study at 86 GHz or more comb spacing [18]. Thus, rigorously speaking, the CMOS-compatible planar microresonator output cannot yet be treated as an optical frequency comb and utilized as an optical frequency calibration standard.

Here we report the first fully stabilized CMOS-compatible chip-scale Kerr frequency comb with a frequency relative inaccuracy down to $2.8 \times 10^{-16}$. The silicon nitride spiral resonator is designed and fabricated to generate a Kerr frequency comb, at 18 GHz native spacing and spanning more than 8 THz over more than 400 comb lines. The comb's two degrees of freedom, one of the comb line frequencies and the comb spacing, are phase locked to a known optical reference and a microwave synthesizer respectively. Active stabilization on the comb spacing improves the RF stability by six orders of magnitude, reaching residual instrument-limited close-to-carrier (10 Hz) phase noise of -70 dBc/Hz and Allan deviation of 3.6 mHz/$\sqrt{\tau}$. In the optical frequency, thirty-three lines of the Kerr frequency comb subset are selected and compared against the current benchmark fiber laser frequency comb and the frequency relative inaccuracy of the stabilized Kerr frequency comb is demonstrated down to 53 mHz. The reported system is a promising compact platform for coherent Raman spectroscopy [19], optical clockwork [20-21], coherent optical communications [22], optical arbitrary waveform generation [23], and astrophysical spectrography [24-26].

Figure 1a shows the experimental setup for the Kerr frequency comb generation and stabilization. The silicon nitride spiral resonator is fabricated with CMOS-compatible processes and the waveguide cross-section is designed to have small and flattened group velocity dispersion for broadband comb generation. Planar ring geometry is employed because of the reduced sensitivity to the environmental perturbation, along with the fewer discrete transverse resonator modes, and the flexibility to tailor the cavity dispersion for efficient and broadband comb generation. The loaded quality factor $Q$ of the pump mode is 660,000 (intrinsic $Q \sim 1,300,000$) and 1 W of pump power is critically coupled to the microresonator, resulting in a maximum coupled pump power 5 times higher than the threshold pump power. Properties of the cold frequency comb are detailed in Supplementary Information Section I. RF amplitude noise and the fundamental beat note are monitored to ensure the Kerr frequency comb is driven from a noisy state to a phase-locked state [17], with details shown in Supplementary Information Section II. The output is first short-pass filtered using a 1550/1590 nm wavelength division multiplexer and then boosted in power with a 13 dBm C-band preamplifier to increase the signal to noise ratio (SNR) of the photodetector signal. Figure 1b shows the Kerr frequency comb spectrum, spanning more than 8 THz and consisting of more than 400 comb lines.

To stabilize the Kerr frequency comb, one of the comb lines and the comb spacing are phase locked to a known optical reference and a microwave synthesizer, respectively. In our system, the known optical reference is derived from an approximately 200 Hz stabilized erbium fiber laser frequency comb (FFC; Menlo Systems) which is also used as a calibration standard later to assess the accuracy of the Kerr frequency comb. A rubidium locked diode laser can also be used as the optical reference [27-28], with details in Supplementary Information Section III. As shown in the Figure 1a, 1% of the pump mode, which is also the strongest Kerr frequency comb line, is tapped and beat with the optical reference on a photodetector. To ensure the beat note has sufficient signal to noise ratio for reliable feedback stabilization (more than 35 dB with 100 kHz RBW), a 0.2 nm bandwidth monochromator is built to filter the FFC before it is beat with the pump. Figure 2a is the free-running beat note, showing a few MHz pump frequency drift in one second. For high bandwidth control of the pump frequency, the diode current of the external-cavity diode laser (ECDL) is directly modulated. Such high bandwidth feedback control, however, has a tradeoff of amplitude modulation of the pump power and consequently excess instability in the comb spacing (Fig. S3). This effect is partly compensated by saturating the erbium doped fiber amplifier and later eliminated by the second feedback loop on the comb spacing. Figure 2b shows the stabilized beat note, illustrating a clear single peak at the center with uncompensated noise above the feedback bandwidth of 300 kHz. The beat has a 70 MHz offset to allow RF amplification for higher signal-to-noise in the feedback loop. Figure 2c is the zoom-in view of the stabilized beat note, showing a resolution limited linewidth of 6 Hz. To quantify the long term stability of the locked pump frequency, the beat signal is analyzed by a frequency counter and the counting results are shown in Figure 2d. The pump frequency remains steady with some occasional noise spikes (peak-to-peak deviation of 17 Hz) and the standard deviation over 10 minutes is as small as 1.2 Hz, limited only by the uncompensated high frequency noise above the feedback bandwidth.

The comb spacing of 17.9 GHz is directly measurable by sending the output to a high speed photodetector (3 dB bandwidth of more than 15 GHz). An 18 GHz local oscillator is used to downmix the electronic signal to the baseband for analysis. The inset of Figure 2e plots the free-running comb spacing beat with a scan range of 1 GHz, showing a clean single peak characteristic of an equidistant Kerr frequency comb. Figure 2e illustrates the non-collinear second-harmonic-generation optical intensity autocorrelation to characterize the temporal structure of the Kerr frequency comb. Careful checks are done to make sure no collinear second-harmonic background is collected in the setup. Even though the Kerr frequency comb is operated in a low noise state, clean circulating mode-locked pulses [9] are not formed as evidenced by the elevated AC background of nearly half of the peak. Furthermore, the autocorrelation measurements are performed at three different delays, showing evidently the repetitive temporal structures of the Kerr frequency comb and excluding the possibility of noise correlation. Here, a fixed phase relationship between different comb lines is obtained, but the phase relationship may contain some abrupt changes associated with the local dispersion disruptions. Thus, mode-locking is prohibited and self-injection locking [11,14] becomes the underlying mechanism for driving the Kerr frequency comb into a low noise state.

The comb spacing is then phase locked and stabilized to a microwave synthesizer by controlling the pump power with a fiber electro-optic modulator (EOM). Pump power is an effective way to control the comb spacing through thermal expansion and thermo-optic effects [29] and nonlinear phase accumulation. Figure 3a shows the stabilized beat note, with a resolution limited linewidth of 6 Hz and a low close-to-carrier phase noise. To characterize the frequency stability of the comb spacing, single sideband (SSB) phase noise spectra and Allan deviations are measured and shown in Figure 3b. Free running, the phase noise of the comb spacing shows a $f^{-3.5}$ dependence on the offset frequency in the vicinity of the carrier. Such close-to-carrier behavior suggests the phase noise is currently dominated by a mixture of technical noise of frequency flicker (30 dB/decade) and frequency random walk (40 dB/decade), rather than limited by quantum noise phase diffusion [30]. Since the microresonator is not thermally insulated from the environment, its interaction with the fluctuating ambient temperature results in the random walk of the comb spacing. Meanwhile, the pump wavelength drift leads to the flicker noise mediated by the residual optical absorption in the microresonator [17]. Such technical noise, however, can be removed by phase locking the beat note to a high performance microwave synthesizer. As shown in Figure 3b, the resulting close-to-carrier phase noise can reach the level of -70 dBc/Hz at 10 Hz with a $f^{-1.5}$ dependence on the offset frequency, limited only by the noise of the microwave synthesizer.

For offset frequency above 10 kHz, the phase noise of the fully stabilized comb spacing is better than that of the 18 GHz local oscillator used for downmixing the electronic signal. The measurement is instrument limited to the level of $\geq -108$ dBc/Hz from 10 kHz to 300 kHz and $-130$ dBc/Hz at 1 MHz. It is therefore informative to calculate the theoretical limit of the phase noise at large offset frequencies and compare with the measurement. Using the equations with the detuning of $\frac{T_R \gamma^2}{D}$ derived in Ref.[30] and assuming $\left(\frac{f}{\gamma}\right)^2 \ll 1$, we obtain the lower limit of the phase noise expressed as

$$\mathcal{L}(f) \approx \frac{2\sqrt{2}\pi\hbar c n_2}{n_0^2 V_0} Q^2 \left[\frac{23}{24} + \left(\frac{4+\pi^2}{96\pi^2}\right)\frac{\gamma^2}{f^2}\right] \quad (1)$$

where $D \equiv (\omega_{m+1} - \omega_m) - (\omega_m - \omega_{m-1})$, $Q$, $n_0$, $n_2$, $V_0$, $2\gamma$, and $f$ are the non-equidistance of the cold cavity modes, quality factor, linear refractive index, nonlinear refractive index, mode volume, FWHM resonance linewidth, and offset frequency respectively. For our spiral microresonator, the estimated phase noise at 1 MHz is $-148$ dBc/Hz and it grows quadratically with the inverse of the offset frequency. The estimated phase noise reaches $-108$ dBc/Hz at 10 kHz and starts to exceed the noise level of the 18 GHz local oscillator, matching the experimental observations. Of note, Eq. (1) derivation requires a single-moded microresonator and the Kerr frequency comb to be mode-locked, and hence only serves as a lower limit to our measurements.

Figure 3b inset plots the Allan deviations of the comb spacing under different conditions. Free running, the Allan deviation increases as $\tau^{1/3}$ as the result of technical noise including the pump wavelength drift and the fluctuating ambient temperature (black open squares). Pump frequency stabilization reduces the increase of Allan deviation over the gate time, but interestingly the level of Allan deviation remains unimproved because of the additional pump power fluctuation from the employed pump frequency control (red semi-open squares). With pump power feedback control, the active stabilization on the comb spacing improves the long-term stability by six orders of magnitude, reaching 3.6 mHz/$\sqrt{\tau}$ (blue closed squares). The residual comb instability is limited by the microwave synthesizer and close to the counter limit at 1 second gate time.

To assess the accuracy of the stabilized Kerr frequency comb, we use the Menlo FFC as the calibration standard and measure the out-of-loop frequencies of thirty-three Kerr frequency comb lines around 1576 nm by beating each comb line with the adjacent FFC mode as shown in the metrology segment of Figure 1a. When the comb spacings of the FFC and Kerr frequency comb are made unequal, the beat frequencies should strictly follow the relationship of

$$f_{beat}^n = \delta + n\left(f_{R,KC} - \left\lfloor\frac{f_{R,KC}}{f_{R,FFC}}\right\rfloor f_{R,FFC}\right) \qquad (2)$$

where $\delta$ is the beat frequency at the pump mode, $f_{R,KC}$ is the Kerr frequency comb spacing, and $f_{R,FFC}$ is the FFC comb spacing. Deviation from this expression is a measure of the frequency inaccuracy of the Kerr frequency comb. Figure 4a shows two example histograms of the frequency counting measurement. 600 counts are accumulated at 1 second gate time for the statistical analysis and the Gaussian curve fitting is implemented to derive the mean values and standard deviations. Counting results on all thirty-three comb lines are shown in Figure 4b. The mean values of the comb frequencies stray from the ideal with a 190 mHz peak-to-peak deviation and a 53 mHz standard deviation. The frequency relative inaccuracy of the stabilized chip-scale frequency comb is thus calculated at 2.8×10$^{-16}$, referenced to the optical carrier at 188 THz.

With the counter frequency error at 10 mHz, we believe the additional measured inaccuracy arises from the third-order dispersion (TOD) of the microresonator. When the TOD is included, the non-equidistance of the cold cavity modes, $D$, can be expressed as:

$$D = -\frac{(\beta_2 L)}{2\pi}\omega_{FSR}^3 + \frac{(\beta_2 L)^2}{4\pi^2}\omega_{FSR}^5 - \frac{(\beta_3 L)}{4\pi}\omega_{FSR}^4 \qquad (3)$$

where $L$ is the cavity's length, $\beta_2$ is the group velocity dispersion (GVD), $\beta_3$ is the TOD, and $\omega_{FSR}$ is the cavity's free spectral range. While the GVD can be compensated by the self-phase modulation in the hot cavity, cancelation of the TOD is less trivial and the remaining TOD can result in the degradation of the comb accuracy, bounded at 1 Hz (assuming no TOD compensation at all) in our microresonator. Of note, the 18 GHz comb spacing directly generated from the microresonator is compatible for high resolution astrospectrography and thus external Fabry-Perot filtering cavities, which limits the precision of state-of-the-art astrocomb [24-26], is circumvented. The 53 mHz frequency inaccuracy of the Kerr frequency comb can potentially improve the precision in astrophysical radial velocity measurements by orders of magnitude.

In summary, we report the first fully stabilized CMOS-compatible chip-scale optical frequency comb. Based on the silicon nitride spiral resonator, a native 18 GHz Kerr frequency comb is generated and its single-sideband phase noise reaches the instrument limited floor of -130 dBc/Hz at 1 MHz offset. The comb's two degrees of freedom, one of the comb line frequencies and the comb spacing, are phase locked to a known optical reference and a microwave synthesizer respectively, reaching an instrument limited residual comb spacing instability of 3.6 mHz/$\sqrt{\tau}$. Thirty-three Kerr frequency comb lines are compared with the current benchmark FFC and the frequency relative inaccuracy of the stabilized Kerr frequency comb is measured down to $2.8\times10^{-16}$. The reported system is a promising compact platform for coherent Raman spectroscopy, high-precision optical clockwork, high-capacity coherent optical communications, optical arbitrary waveform generation, and astrophysical spectrography.

**Methods**

**Microresonator characteristics**: The silicon nitride waveguide cross-section is designed to be 2 μm × 0.75 μm so that not only the group velocity dispersion but also the third order dispersion is small at the pump wavelength. The spiral design ensures the total footprint of the relatively large resonator can be minimized (less than $0.9 \times 0.8$ mm$^2$), eliminating the additional cavity losses associated with the photomask stitching and discretization errors. The intrinsic quality factor of the spiral resonator is estimated to be 1,300,000. Bends in the resonator have diameters greater than 160 μm to minimize the bending-induced dispersion. Adiabatic mode converters are implemented on the side of the chip to improve the coupling efficiency from the free space to the bus waveguide, to less than 3 dB coupling loss per facet. The gap between the bus waveguide and the microresonator is 500 nm, leading to a critical coupling at the pump wavelength.

**Kerr comb generation**: A tunable external-cavity diode laser (ECDL) is amplified by an L-band erbium doped fiber amplifier (EDFA) to 2W and then coupled to the microresonator. A 1583-nm longpass filter removes the amplified spontaneous emission noise from the EDFA. The microresonator chip temperature is actively stabilized to +/- 10 mK. A 3-paddle fiber polarization controller and a polarization beam splitter cube are used to ensure proper coupling of the TE polarization into the microresonator. To obtain the Kerr frequency comb, the pump wavelength is first tuned into the resonance from the high frequency side at a step of 10 pm until primary comb lines are observed on the optical spectrum analyzer, prior to fine control to drive the comb from a noisy state to a phase-locked state.

**Device fabrication**: First a 3 μm thick SiO$_2$ layer is deposited via plasma-enhanced chemical vapor deposition (PECVD) on *p*-type 8" silicon wafers to serve as the under-cladding oxide. Then low-pressure chemical vapor deposition (LPCVD) is used to deposit a 750 nm silicon nitride for the spiral resonators, with a gas mixture of SiH$_2$Cl$_2$ and NH$_3$. The resulting silicon nitride layer is patterned by optimized 248 nm deep-ultraviolet lithography and etched down to the buried SiO$_2$ via optimized reactive ion dry etching. The sidewalls are observed under SEM with an etch verticality of 82° to 88° (see Supplementary Information Section I). Then the silicon nitride spiral resonators are annealed at 1200$^\circ$C to reduce the N-H overtones absorption at the shorter wavelengths. Finally the silicon nitride spiral resonators were over-cladded with a 3 μm thick SiO$_2$ layer, deposited initially with LPCVD (500 nm) and then with plasma-enhanced chemical vapor deposition (2500 nm). The propagation loss of the $Si_3N_4$ waveguide is ~ 0.2 dB/cm at the pump wavelength.

**Acknowledgements:** The authors thank discussions with Michael McDonald, Heng Zhou, Jinkang Lim, and Abhinav Kumar Vinod. We also acknowledge assistance from Jaime Gonzalo Flor Flores and Yongjun Huang on the cross-section scanning electron micrographs, and loan of the microwave signal generator from the Bergman group at Columbia University. The authors acknowledge funding support from DARPA (HR0011-15-2-0014), NIST Precision Measurement Grant (T. Zelevinsky and B. H. McGuyer; 60NANB13D163), ONR (N00014-14-1-0041), and AFOSR Young Investigator Award (S. W. Huang; FA9550-15-1-0081).

**Author contributions:** S.W.H. designed the measurements, analyzed the data and wrote the paper. S.W.H and J.H.Y performed the measurements, and S.W.H., J.H.Y., and C.W.W. designed the layout. M.Y. and D.L.K. performed the device nanofabrication. B.M. and T.Z. aided in the measurements with the Menlo fiber laser frequency comb. All authors contributed to the discussion and revision of the manuscript.

**Additional information:** The authors declare no competing financial interests. Reprints and permission information is available online at http://www.nature.com/reprints/. Correspondence and requests for materials should be addressed to S.W.H and C.W.W.

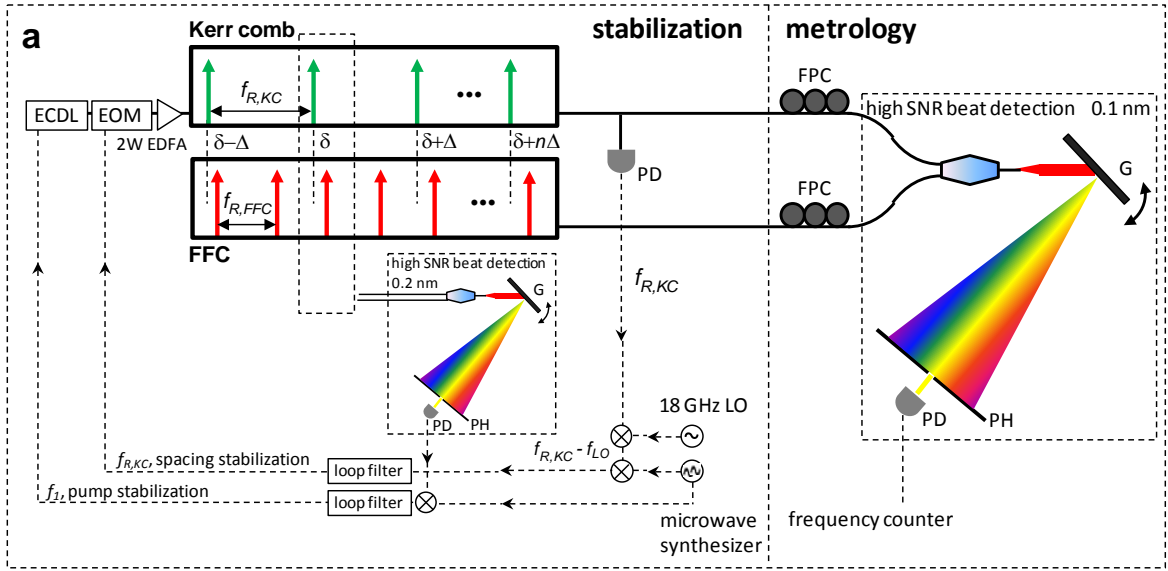

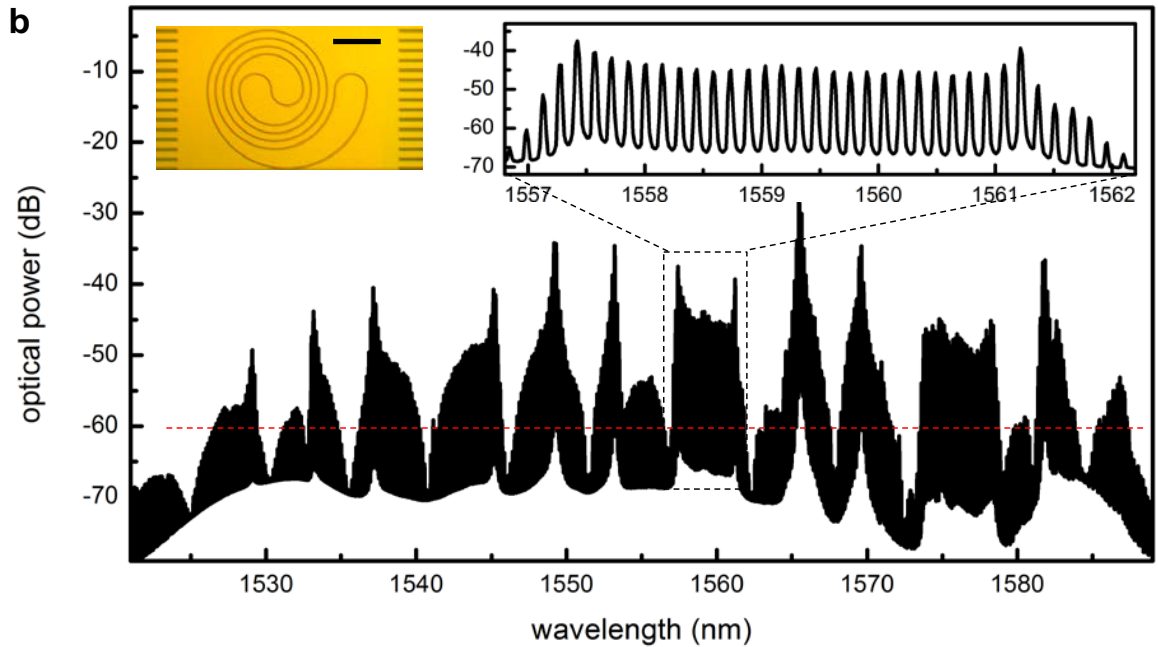

**Figure 1 | A stabilized chip-scale optical frequency comb. a,** Measurement setup schematic for the generation and stabilization of the chip-scale optical frequency comb. To stabilize the comb's first degree-of-freedom, an ECDL is phase-locked to an optical reference, here a mode of a stabilized fiber laser frequency comb, and then amplified to 2 W to drive the $Si_3N_4$ microresonator. To stabilize the comb's second degree-of-freedom, the Kerr comb spacing, $f_{R,KC}$, is monitored by sending the comb to a high-speed photodetector (3 dB bandwidth more than 15 GHz) and downmixing the electronic signal to the baseband with a local oscillator at $f_{LO} = 18$ GHz. A fiber electro-optic modulator (EOM) controls the pump power and stabilizes the comb

spacing via thermal effect. δ: frequency offset between the pump and the adjacent fiber laser frequency comb line. Δ: comb spacing difference between the Kerr comb and the fiber laser frequency comb. In the comb stabilization segment, PD: the photodetector; G: grating; and PH: pinhole. Dashed lines denote the electronic loops, and solid lines denote the optical paths. In the comb metrology segment, FPC denotes the fiber polarization controller. **b,** Example stabilized Kerr frequency comb, consisting of more than 400 comb lines in the telecommunication wavelength range. The horizontal (red) dashed line denotes the 1 µW per comb line power level. Left inset: optical micrograph of the spiral microresonator. Right inset: comb lines with native spacing at the cavity's free-spectral range are clearly observed.

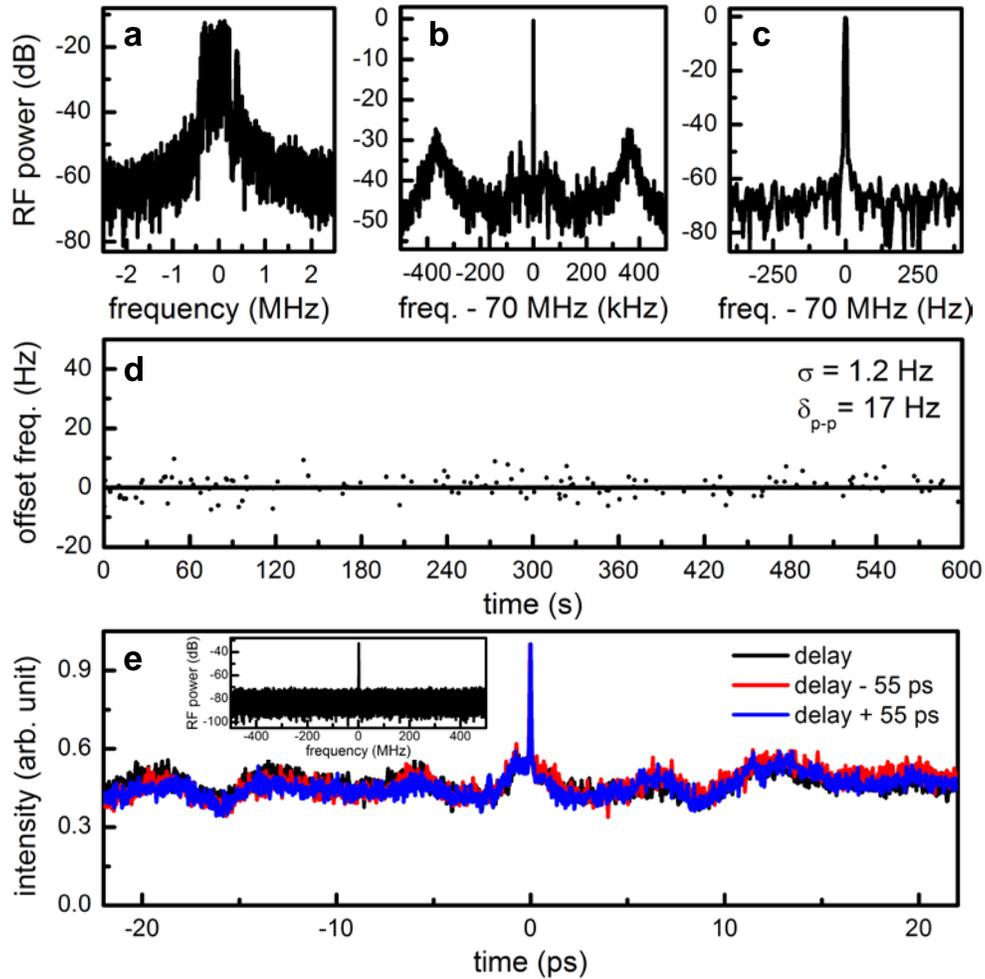

**Figure 2 | Stabilizing the pump frequency to the Hz level residual error and temporal structure of the phase-locked Kerr comb. a,** Free-running beat note between the pump and the fiber laser frequency comb. To obtain a sufficient signal-to-noise ratio for reliable feedback stabilization (more than 35 dB with 100 kHz RBW), a 200 pm bandwidth monochromator is built to filter the fiber frequency comb before it is mixed with the pump. **b,** RF spectrum of the stabilized beat note. Control of the pump frequency is achieved by modulating the ECDL diode current, with 300 kHz bandwidth. **c,** Zoom-in view of the stabilized beat note, showing a resolution limited linewidth of 6 Hz. **d,** Frequency counting of the stabilized beat note with a gate time of 1 s. The standard deviation over 10 minutes is 1.2 Hz with some occasional noise spikes, limited by the uncompensated high frequency noise above the ECDL diode current feedback bandwidth. **e,** Optical intensity autocorrelations of the phase-locked Kerr frequency comb at different delays, showing evidently the repetitive structures and excluding the possibility of noise correlation. Inset: RF spectrum of the free-running comb spacing with a scan range much larger than the cavity linewidth (290 MHz). The comb is tuned to be offset-free by fine control of the pump frequency.

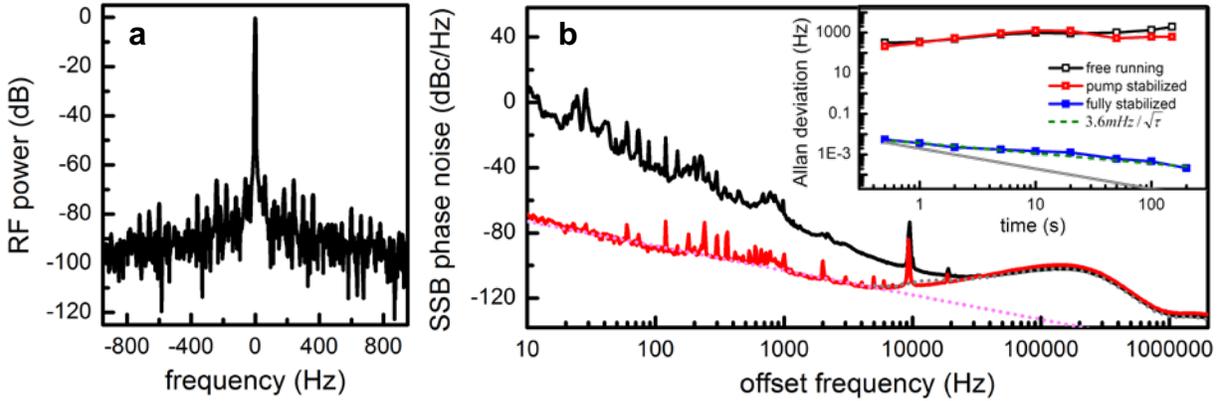

**Figure 3 | Stabilizing the comb spacing to the mHz level residual error. a,** RF spectrum of the stabilized comb spacing, showing a resolution limited linewidth of 6 Hz. Control of the comb spacing is achieved by modulating the pump power via a fiber EOM. **b,** Single-sideband (SSB) phase noises of the free-running (black curve) and stabilized (red curve) comb spacing. Free running, the phase noise of the comb spacing shows a $f^{-3.5}$ dependence on the offset frequency in the vicinity of the carrier. Such technical noises can be removed by phase locking the beat note to a high performance microwave synthesizer and the resulting close-to-carrier phase noise can reach the level of -70 dBc/Hz at 10 Hz with a $f^{-1.5}$ dependence on the offset frequency (pink dashed curve), limited only by the microwave synthesizer. For offset frequencies above 20 kHz, on the other hand, the phase noise of the comb spacing is better than that of the 18 GHz local oscillator used for downmixing the electronic signal (gray dashed curve) and the measurement is thus instrument limited. Inset: Allan deviation of the comb spacing under free-running (black open squares), pump frequency stabilization (red semi-open squares) and full stabilization (blue closed squares). The stabilized comb spacing shows a consistent trend of 3.6 mHz/$\sqrt{\tau}$ (green dashed line) when the gate time is in the range from 0.5 s to 200 s. The gray line denotes the counter-limited Allan deviation.

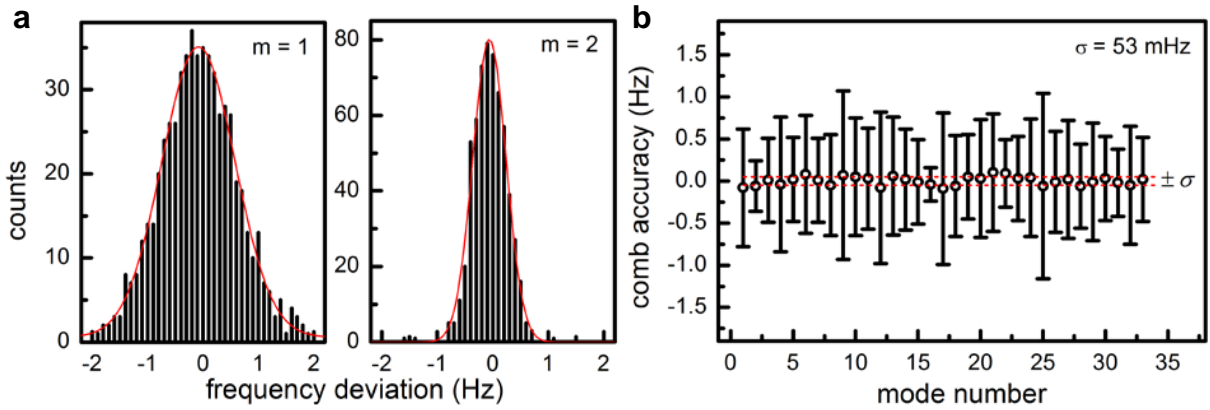

**Figure 4 | Out-of-loop characterization of the fully-stabilized chip-scale optical frequency comb.** To quantify the accuracy of the stabilized chip-scale optical frequency comb, each of the comb lines around 1576 nm ($m$=16) is mixed with the fiber laser frequency comb and the beat frequency is counted with a gate time of 1 second. The beat frequencies should change progressively by $\Delta$, where $\Delta = f_{R,KC} - Nf_{R,FFC}$, and the deviation from this relationship is a measure of the accuracy of the chip-scale optical frequency comb. **a,** Example histograms of the frequency counting measurement on the 1$^{st}$ and 2$^{nd}$ modes ($m$). 600 counts are accumulated for the statistical analysis. The red lines are the Gaussian fits to the histograms. **b,** Counting results on the optical frequencies of 33 comb lines. The centroid of the comb frequencies stray from the ideal with a 190 mHz peak-to-peak deviation and a 53 mHz standard deviation. The frequency relative inaccuracy of the stabilized chip-scale frequency comb is thus calculated at $2.8\times10^{-16}$, referenced to the 188 THz optical carrier.

# Supplemental Information for

# A stabilized 18 GHz chip-scale optical frequency comb at 2.8×10<sup>-16</sup> relative inaccuracy

S.-W. Huang[1,*], J. Yang[1], M. Yu[2], B. H. McGuyer[3], D.-L. Kwong[2], T. Zelevinsky[3], and C. W. Wong[1,*]

[1] *Mesoscopic Optics and Quantum Electronics Laboratory, University of California, Los Angeles, CA 90095*

[2] *Institute of Microelectronics, Singapore, Singapore 117685*

[3] *Department of Physics, Columbia University, New York, NY 10027*

## I. Properties of a fully-stabilized chip-scale Kerr frequency comb

Figure S1a shows a cross-section scanning electron micrograph of the microresonator waveguide, with an estimated 82° to 88° slope of the vertical sidewalls. The refractive index of the low pressure chemical vapor deposition (LPCVD) $Si_3N_4$ film was measured with an ellipsometric spectroscopy (Woollam M-2000 ellipsometer) and then fitted with the Sellmeier equation assuming a single absorption resonance in the ultraviolet. The fitted Sellmeier equation, $n(\lambda) = \sqrt{1 + \frac{2.90665\lambda^2}{\lambda^2 - 145.05007^2}}$, and the measured 88° sidewall angle were both imported into the COMSOL Multiphysics for the microresonator design. Figure S1b shows the modeled free spectral range of the first two TE modes of the microresonator. While the fundamental mode features a FSR of 17.9 GHz, the $TE_2$ mode has a slightly lower FSR and thus the resonances of the $TE_2$ family approaches that of the fundamental family about every 4 nm ($\frac{FSR^2}{\Delta FSR} = 460 GHz$). The mode interaction when the resonances are close leads to local disruption of the phase matching condition [S1-S3] and results in the periodic amplitude modulation on the Kerr comb spectrum (Fig. 1b).

Due to the large refractive index of the $Si_3N_4$ waveguide, a 600 μm long adiabatic mode converter (the $Si_3N_4$ waveguide, embedded in the 5×5 μm² $SiO_2$ waveguide, has gradually changing widths from 0.2 μm to 1 μm) is implemented to improve the coupling efficiency from the free space to the bus waveguide. The input-output insertion loss for the waveguide does not exceed 6 dB.

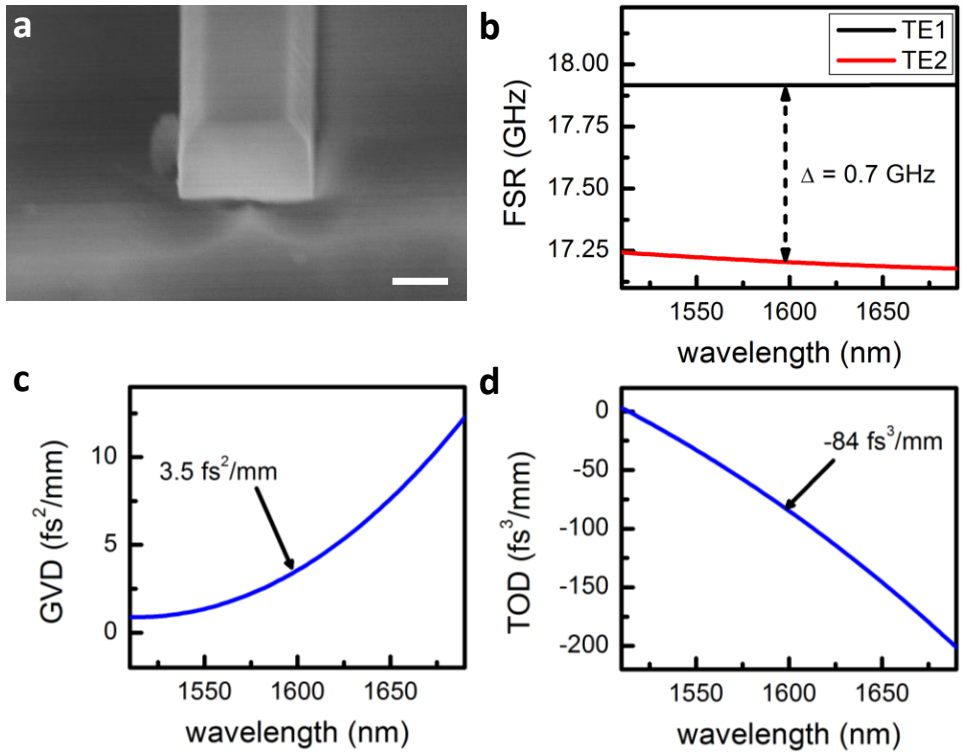

**Figure S1 | Properties of the fully stabilized 18 GHz Kerr frequency comb. a,** Scanning electron micrograph of the waveguide cross-section. Scale bar: 500 nm. **b,** Modeled free spectral range of the first two TE modes of the chip-scale optical frequency comb. **c,** Modeled group velocity dispersion of the fundamental mode, measuring a GVD of 3.5 fs$^2$/mm at the pump wavelength. **d,** Modeled third order dispersion of the fundamental mode, measuring a TOD of -84 fs$^2$/mm at the pump wavelength.

## II. Low-noise state of the Kerr frequency comb

As the pump wavelength was tuned into the resonance from the high frequency side, we first observed multiple RF spikes because the primary comb line spacing is incommensurate with the fundamental comb spacing. The state with incommensurate spacing was unstable and it made frequent transition to high-noise state characterized by elevated RF amplitude noise (45 dB higher than the phase-locked comb state). Next, with fine control of the pump wavelength (10 MHz/step), the offset between different comb families can be made zero such that the RF amplitude noise spectrum showed no excess noise (Figure S2). The phase-locked comb typically stabilized for hours.

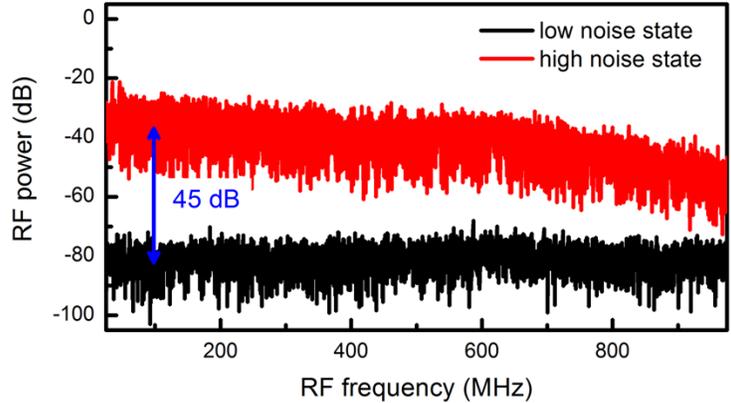

**Figure S2 | RF amplitude noise spectra of the high noise state and the low noise phase-locked comb state.** With the proper pump wavelength, RF amplitude noise dropped by 45 dB and approached the detector background noise, indicative of the transition into a phase-locked state.

### III. *After*-resonator feedback stabilization scheme and measurements

As discussed in the main text, a rubidium locked diode laser can also be used as the optical reference for phase locking one of the comb lines [S4-S6]. In the alternative scheme, the comb line has to be tapped after the microresonator and thus the prerequisite for such scheme is the ability to phase lock the pump after the comb generation stage (Figure S4). Instead of tapping the pump light at the very beginning, here the pump was split and beat with the optical reference on a photodetector after the comb generation stage. The rest of the setup was the same as the one shown in Figure 1a. Figure S4b shows that the beat note can be equally well stabilized to a resolution limited linewidth of 6 Hz.

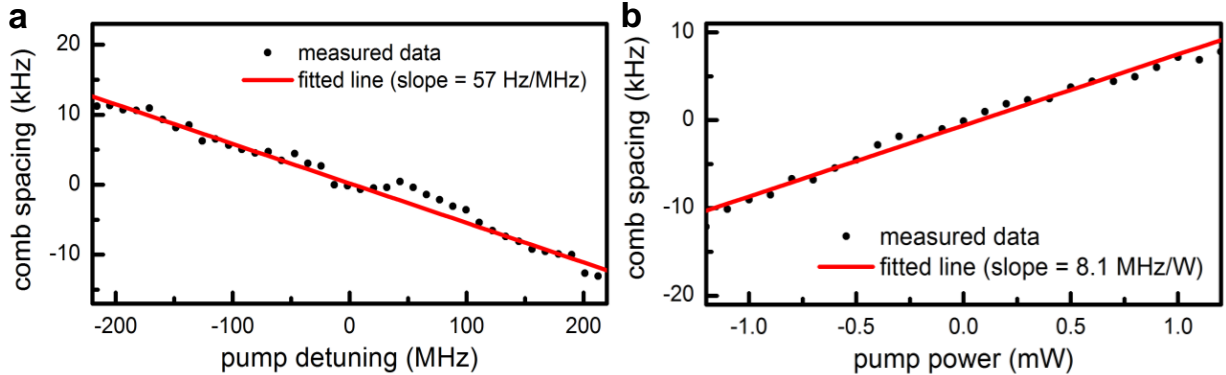

**Figure S3 | Dependence of the comb spacing on the pump properties. a,** The comb spacing as a function of the pump wavelength detuning, determined at 57 Hz/MHz in our microresonator. **b,** The comb spacing as a function of the pump power in the ring, determined at 8.1 MHz/W in our microresonator.

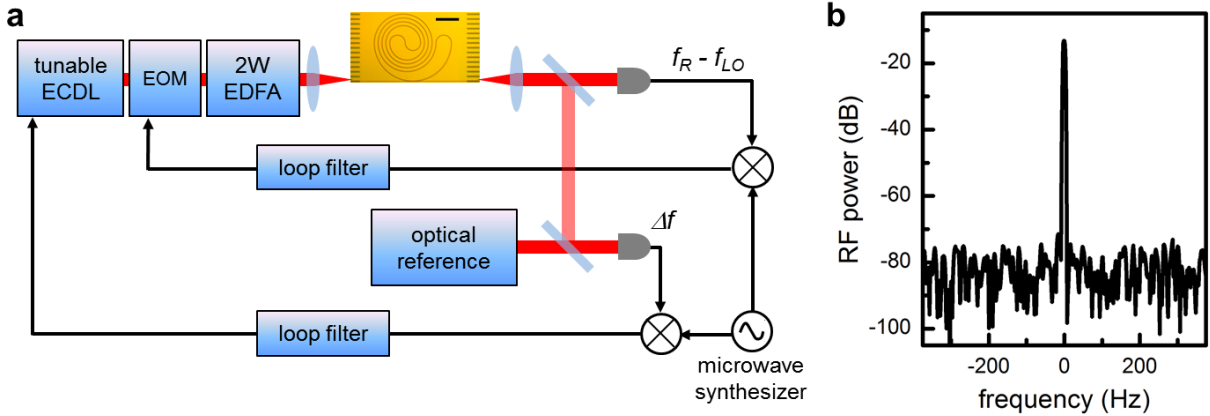

**Figure S4 | Schematic of the alternative experimental setup for generation and stabilization of the chip-scale optical frequency comb. a,** Instead of tapping the pump light at the very beginning, here the pump was split and beat with the optical reference on a photodetector after the comb generation stage. The rest of the setup was the same as the one shown in Figure 1a. **b,** RF spectrum of the stabilized beat note, showing a resolution limited linewidth of 6 Hz. Control of the pump frequency was again achieved by modulating the diode current of the ECDL.